# Extraordinary optical transmission through metal films with sub wavelength holes and slits

A. S. Vengurlekar

Tata Institute of Fundamental Research, Mumbai 400005, email asv@tifr.res.in

**Abstract:**

Continuous films of metals like gold and silver with a thickness of a few tens of nm have poor optical transmission in the visible and infrared. However, the same films become largely transparent when the transmission is mediated by coupled surface plasmon polaritons on the two surfaces of the film. Likewise, it is expected that optical transmission through a hole in an opaque metal film would be negligible if the size of the hole is much smaller than the wavelength of the incident radiation. Contrary to this, it was found a few years ago that opaque metal films perforated with a periodic array of sub wavelength holes exhibit large transmission at certain wavelengths. Such extraordinary optical properties of the metal films have attracted much attention in recent years. They have led to several proposals and demonstrations of nanophotonic applications in a wide range of areas, such as microscopy, spectroscopy, optoelectronics, optical data storage, bio-chemical sensing and so on. In parallel, extensive effort has been made to elucidate the mechanisms of efficient light energy transfer across metal films with arrays of sub wavelength apertures like holes and slits. Despite this effort, not all aspects of the underlying physics of the phenomena are fully understood yet. Here, we provide a short review of the various concepts put forth to explicate the large transmission of light across continuous metal films and films with sub wavelength apertures.

**I Continuous metal films:**

It is well known[1,2] that electromagnetic plane waves incident from free space on a metal become mostly inhomogeneous and non propagating inside the metal, unless their frequency exceeds the bulk plasmon frequency. In general, the wave fronts of constant amplitude and phase then are not parallel. The amplitude of the wave decays exponentially inside the metal with a decay constant $\delta=2\pi\kappa/\lambda$ where $\kappa=\text{Im}\{\eta\}$, $\eta=$ refractive index of the metal, $\lambda =$ wavelength in free space. The transmittance ($=I/I_0$, I and $I_0$ are incident and transmitted power of radiation) of a metal film with thickness h of a few tens of nms is usually rather small at visible wavelengths. For example, since $\kappa\cong 4$ for silver[3,4] at 600nm, the transmittance is only about 1 percent for h=50nm. However, the same semi-opaque film can become largely transparent at visible wavelengths under certain conditions, as described below.

Interfaces of a dielectric with metals like gold and silver can support propagating transverse magnetic (TM) modes of electromagnetic field that are bound to the interface and are coupled to collective oscillations of electron density at the metal surface. The field of these modes is enhanced on the interface, but decays in both directions normal to the interface on the scale of several nms. These are the well known surface plasmon polariton (SPP) modes[5,6]. They can occur on metal-dielectric interfaces in a frequency range where $\text{Re}\{\varepsilon_m\}<0$, $|\text{Re}\{\varepsilon_m\}|> \varepsilon_d$ is possible. Here $\varepsilon_d$ and $\varepsilon_m$ are $\omega$ dependent permittivities of the dielectric and metal respectively, relative to vacuum. The in-plane SPP dispersion is given by $k = (\omega/c)[\varepsilon_d\varepsilon_m/(\varepsilon_d+\varepsilon_m)]^{1/2}$, where $\omega=$angular frequency of light. The Drude form: $\varepsilon_m = 1-\omega_p^2/[\omega(\omega+i\omega_\tau)]$ for a free electron metal is often used as an approximation, where $\omega_p$ ($=[4\pi Ne^2/m_0]^{1/2}$, N= electron density, $m_0$=electron effective mass) and $\omega_\tau$ are respectively the bulk plasma and collision frequencies. For very large k, the SPP dispersion approaches the surface plasmon frequency $\omega_{SP}$, unique for a given metal and dielectric and given as $\omega_p/(1+\varepsilon_d)^{1/2}$ when $\omega_{SP} > \omega_\tau$. (For silver-air interface, for example, $\varepsilon_d=1$, and $\omega_{SP} \cong 4\times 10^{15} sec^{-1}$ and $\omega_\tau \cong 3\times 10^{13} sec^{-1}$ for silver[4]). Surfaces of gold and silver in air can support SPPs in a broad range of wavelengths, including the visible and infrared (IR). Monochromatic plane waves of light incident from a dielectric on top of the metal can not couple to the SPP at a smooth metal-dielectric interface at any angle of incidence ($\theta$). The reason is that the wave vector ($=k_{SP}$) of an SPP propagating along the interface can not match the projection $nk_0 \sin\theta$ of the wave vector of incident light on the metal-dielectric interface at any $\theta$ (Fig. 1). Here n=refractive index of the dielectric and $k_0 = \omega/c = 2\pi/\lambda$. The largest projection of the light wave vector ($=nk_0$) on the metal-dielectric interface is still smaller than $k_{SP}$. For the SPPs to be optically excited, two conditions need to be satisfied: first, the light has to be TM i.e. p-polarized, with the electric field vector in the plane of incidence, and second, the phase matching condition has to be satisfied. In certain situations, these conditions become unimportant. For example, SPP excitation by light, irrespective of its polarization, is possible when the metal surface has roughness or spatially localized scattering centers such as nano-dents, -holes or -bumps that provide the missing momentum needed for phase matching between incident light and SPPs[5]. Of course, correspondingly, the SPP scattering and radiative loss would be larger on a rough interface. In another instance, a focused light beam is known to excite SPPs even on smooth films[7].

Consider an interface between two dielectric media, say $D_{II}$ and $D_I$ of refractive indices $n_{II}$ and $n_I$ respectively, with $n_{II} > n_I$. A monochromatic wave incident from $D_{II}$ toward the interface at an angle of incidence $\theta$ undergoes total internal reflection (TIR) at the interface at $\theta = \theta_{TIR}$, with $\theta_{TIR}$ given by $\sin(\theta_{TIR}) = n_I/n_{II}$. The wave refracted in $D_I$ at the interface then propagates along the interface. For $\theta > \theta_{TIR}$, it becomes evanescent with its field decaying inside $D_I$ (Fig. 2a). Since it has a propagation constant $n_{II}k_0\sin\theta$, it is possible for a certain choice of $n_{II}$, $n_I$ and $\theta$ to obtain $n_{II}k_0\sin\theta = k_{SP}$ ($>n_Ik_0$) where $k_{SP}$ is the wave vector of an SPP on the metal-$D_I$ interface. This is the basis of two well known schemes[5] of phase matching of plane waves of light with SPPs on a smooth interface of a dielectric with metal. In one, called the Kretschmann geometry (Fig.2b), a 3 layer stack consisting of $D_{II}$–metal-$D_I$ is used. If the metal is in the form of a thin film, with its thickness not much larger than the skin depth (e.g. about 25nm for gold at 600nm), the evanescent wave described above tunnels across it to excite an SPP at the metal-DI interface. The other scheme uses the so called Otto configuration which typically has a 3 layer stacked structure of $D_{II}$ - $D_I$ – metal (Fig.2c). The tail of the evanescent wave refracted along the $D_{II}$ - $D_I$ interface and decaying inside $D_I$ can reach the $D_I$ -metal interface if $D_I$ is thin enough (thickness not much larger than the wavelength). If the light is p-polarized, it can couple to the SPP at the $D_I$ -metal interface at a wavelength that is determined by $\theta$.

If the metal film in the Otto geometry has the same dielectric ($D_I$) on the transmission side, the SPPs at the two $D_I$ -metal interfaces of the symmetric structure: $D_I$ -metal -$D_I$ are degenerate. They are uncoupled[8,9] if the metal film thickness (h) is large enough (h>> $n_Ic/\omega_p \sim$ 70nm for Ag covered by MgF$_2$, for example, with $n_I$ =1.38 for MgF$_2$[10], $\omega_p \cong$ 6x10$^{15}$sec$^{-1}$ for Ag[4]). For smaller h, the SPPs on the two interfaces get coupled and split into two modes– one mode with its magnetic field in the plane of the film symmetric with respect to the two interfaces and the other asymmetric with a node in the metal[9]. The former gets more damped in the metal. The evanescent wave at the $D_{II}$-$D_I$ interface beyond the TIR condition can resonantly excite the coupled SPP modes of the structure: $D_I$ -metal - $D_I$. If the transmission side dielectric ($D_I$) is supported by another layer of $D_{II}$ (so that one has a completely symmetric structure $D_{II}$ - $D_I$ -Metal -$D_I$ - $D_{II}$ as in Fig.3), the excited SPP modes can get reconverted into a propagating wave in $D_{II}$ on the transmission side. The enhanced fields at the interfaces due to the SPPs play an important role in causing large transmission. To sum up, the incident light first excites the coupled SPP modes on both sides of the metal film, and those modes then re-radiate as transmitted beam. If absorption in the metal film is small, it is possible in the above scheme to have a large enhancement in transmittance of an otherwise semi opaque metal film. Dragila et al[10] demonstrated that transmittance of a 60nm silver film at 632.8nm (HeNe laser line) with $\theta \cong 74^0$ is about 75% when embedded in a $SiO_2$-$MgF_2$-Ag-$MgF_2$-$SiO_2$ symmetric structure, an enhancement of several orders over the less than 1% transmittance normally obtained for a free standing Ag film of the same thickness. One can imagine an angle tunable, efficient color selective transmission filter using this.

Use of the multilayer dielectric structure described above is not the only way to achieve SPP mediated enhanced light transmission across thin metal films. Another way is to use a metal-dielectric interface that has a one-dimensional periodic corrugation perpendicular to the plane of incidence[5] (Fig.4). The phase matching condition for p-polarized light incident from a dielectric on top of the metal surface and the SPP on the metal-dielectric interface then becomes: $k_{SP} = nk_0\sin\theta \pm mG$, $G=2\pi/\Lambda$, $\Lambda=$ the grating period and m=1,2,3…. In general, the dispersion of the SPP Bloch mode excited on a grating may deviate from that on a flat interface. The change is often disregarded for small modulation (except at points where different branches of the SPP dispersion cross and interact). Let us consider the metal to be in the form of a film with both interfaces identically corrugated. If the dielectrics on its two sides have the same n, the SPPs on the two interfaces become degenerate. If the metal film is thin enough (a few tens of nms), the SPPs split into two coupled modes[11-13]. Once again, efficient transmission of incident light across the metal film is possible when the light resonantly excites the coupled SPP modes. If the dielectrics on the two sides of the thin metal film have very different refractive indices (say n and n′), the SPPs on the two interfaces are non degenerate. Light transmission may be still possible across such films. For that, the light first excites SPP on the interface on the incidence side. That SPP can transfer its energy to the SPP on the other interface if phase matching between the two SPPs becomes possible[14]. Finally, the SPP on the transmission side is diffracted by the second grating into transmitted light at an angle $\theta_t$ determined by $n\sin\theta = n′\sin\theta_t \pm m(\lambda/\Lambda)$. It was demonstrated[13] that cross coupling of SPPs on the two sides of a corrugated metal film placed on top of a dielectric fluorescent layer in a light emitting device permits very efficient extraction of light from the active region into free space, with the additional advantage that the emission is directional. The h dependence of the diffracted light intensity usually shows a peak at h of a few tens of nms[15]. The peak occurs because the transmission is essentially determined by a trade off between two opposite effects – on the one hand, the SPP field at the interface on the incidence side increases with h thereby enhancing and on the other hand, damping of the field in the metal becomes larger with h.

Large transmission is possible even when the gratings on the two sides are zero order ($\lambda>2\Lambda$) so that there are no higher diffracted orders. The incident light then can not excite propagating SPPs. Tan et al[16] studied silver films with narrow grooves on the two sides and showed that high transmission was nevertheless possible because of tunneling of light via coupled plasmon modes localized in the grooves. Smooth, uncorrugated thin metal films with a dielectric grating on top can also show large and directional transmission due to light-SPP coupling via the dielectric grating[17,18].

The special schemes of multiple layered metallo-dielectric structures or corrugated metal films described above are not necessary for energy transfer across a metal film when the SPPs on the two interfaces can couple to the near fields of excited dipoles in close vicinity of the metal surfaces. This kind of mechanism is quite efficient for molecular donor-acceptor energy transfer and can operate even for metal thickness upto several tens of nms[19], which is too large for the well known Dexter and Forster mechanisms to be active. We may also mention that a thin film of a metal like silver can act as a semi-transparent superlens[20,21], focusing near fields of a dipole emitter with a resolution that is far better than that set by the diffraction limit for propagating far fields. The key feature here is that unlike the far fields, the near fields have large transverse wave vector components that contain information on the sharp spatial features of the emitter. The near fields of the dipole couple to the SPPs, the evanescent fields of which then compose the image on the other side of the film.

**II Sub wavelength holes in an opaque metal film:**

Continuous metal films become essentially opaque to light transmission if the film thickness h is large (e.g., h>100nm for silver in the visible). It is clear that when an optically thick metal film has a hole, transmission of light, if any, can occur only through the aperture. How much of the light falling on a single *subwavelength* size hole in the opaque film gets transmitted? According to the calculations performed by Bethe[22] many years ago, the transmittance of light incident on a circular hole of diameter d is proportional to $(d/\lambda)^4$ when d<< $\lambda$, and therefore is very small. Recently, Ebbesen and co-workers[23,24] measured the light transmitted at visible-infrared (IR) wavelengths through a periodic two-dimensional array of subwavelength holes drilled in an opaque silver film. It came as a big surprise when their results showed that, at certain wavelengths, the transmittance per hole in the array exceeded that expected from Bethe's theory for an isolated hole by several orders. Being able to squeeze light through subwavelength metallic apertures with such high efficiency has many implications for novel applications of nanophotonics[25-27] in microscopy, spectroscopy, optoelectronics, optical data storage and bio-chemical sensing. The Ebbesen result therefore aroused tremendous excitement. The 'extraordinary transmission' has since remained a topic of intense research and some debate in the literature, with several hundreds of papers published. These investigations have now led to a reappraisal of the actual size of the enhancement[28-30]. It is realized that transmission through a hole placed in the array should be compared with that in a single isolated hole prepared under identical experimental conditions (Fig.5), and not with Bethe's theory whose validity regime is not appropriate for the Ebbesen experiments. Under conditions typically used in such experiments, light transmission for $\lambda > d$ in a single isolated hole actually turns out to be much larger than that suggested by Bethe's theory. Therefore the claim of a hugely enhanced transmission across a hole in the array relative to that an isolated hole can be misleading. Nonetheless, it remains a fact that the transmission is indeed enhanced, by about one order or so, if not several. What is more, at certain wavelengths, the total photon flux emerging from a hole in an array actually exceeds that incident on it. Since energy conservation can not be violated, it is clear that light falling on regions surrounding a hole in the array must be somehow getting funneled into a hole so that, in effect, the hole collects light incident over an area larger than its own. Another feature of great interest is that, in contrast with the smooth wavelength spectrum of transmission expected for an isolated hole, the spectrum for the hole array shows both enhancement and suppression of light transmission at certain wavelengths.

As for the underlying mechanisms of these 'extraordinary' phenomena, universal consensus is still emerging. Two main points of view have been discussed in the literature. On one hand, many workers accepted that the maxima occur when SPPs on the two interfaces get resonantly coupled via evanescent waves in the hole[31-34]. This picture is rather appealing, being analogous to the case of coupled SPP assisted transmission across thin continuous films described earlier, except that the coupling is now considered to occur via fields tunneling through the holes. On the other hand, there are some who suggested that the interface SPPs do not play any special role; the enhancement arises because of a combined effect of the apertures in the array, such as interference and diffraction of the surface waves excited by light incident on the perforated metal surface[28,30,35,36]. Currently, it seems that the two views are converging to accept that both SPPs and diffraction-interference effects play a role together. These aspects are described in more detail below.

*Single isolated hole*

It is well known[2] that when a parallel beam of light is incident normally on an opaque screen having a small circular hole, the transmitted light diffracts in the half space on the exit side. For large-size holes with d > λ, the transmitted light viewed at a large distance (>>d) shows diffraction rings such that the central bright cone subtends an angle φ with respect to the normal to the screen, given by sinφ ≈ 1.22λ/d. The ratio of the power of transmitted light and of light incident on the hole is of the order of unity and wavelength independent (with d > λ). However when the hole is of subwavelength size, the transmission could become smaller by several orders according to Bethe. Bethe's theory makes the approximation of a very small hole (λ>>d) in a thin film of a *perfect conductor* (i.e. having electrical conductivity $\sigma(\omega) \to \infty$, dielectric function $\varepsilon \to -\infty$, skin depth $\to$ 0) with vanishing thickness. However, these approximations are not suitable for conditions used in Ebbesen's experiments where the measurements are made in the visible and IR, typically on a periodic array of holes in Ag films with d=150nm, h=200nm and Λ=600nm with 8> (λ/d) >2. It therefore would be appropriate to compare Ebbesen's results with theories that go beyond Bethe's approximations, by considering practical situations where h and d of the holes are finite and the metal is not an ideal conductor.

A circular hole in a metal slab is a cylindrical waveguide. Light propagation through the aperture occurs via waveguide modes. Transmission of propagating modes in a very long (h>>d) cylindrical waveguide in a *perfect conductor* has a cut-off wavelength given respectively by the $l^{th}$ root of the $m^{th}$ Bessel function of the first kind and its derivative for the $TM_{lm}$ and $TE_{lm}$ (transverse electric) modes[37]. The hole can support only evanescent light modes for $\lambda > \lambda_c$, where $\lambda_c \sim 1.7 n_0 d$ is the largest cut-off wavelength ($n_0$=refractive index of the dielectric in the waveguide). It occurs for the $TE_{11}$ mode. The transmission in the ideal waveguide decays exponentially[38] for $\lambda > \lambda_c$, faster than Bethe's theory. For realistic conditions, namely finite h and finite $Re\{\sigma(\omega)\}$ and $Im\{\sigma(\omega)\}$, the waveguide transmission gets modified in the following way. When h is finite, comparable to d, numerical calculations[30] show that the transmission for $\lambda > \lambda_c$ does not decay as rapidly as in the long, ideal waveguide case. Further, for realistic metals, the fields have nonzero depth of penetration inside the metal, for both plasmonic metals like Ag and dielectric type metals like tungsten (that have $Re\{\varepsilon\} > 0$ in much of the IR-visible frequency regime). This can make the effective hole area larger and give larger transmission. Finally, for a plasmonic metal like silver with $Re\{\varepsilon\} < 0$ in the visible-IR, new channels of transmission become possible for $\lambda > \lambda_c$. Light incident on the hole can excite a localized surface plasmon (LSP) on the rim of the hole in the form of oscillating magnetic dipoles[29,39]. The associated enhanced evanescent field can tunnel to the exit side of the hole if h is small enough. That excites a corresponding LSP on the exit side rim, which in turn couples to far field radiation giving rise to a broad resonant peak in transmission before the final decay at large $\lambda$. A second possibility is that the internal surface of the hole may support cylindrical SPP modes[30,38] propagating along the hole for certain $\lambda > \lambda_c$. The wavelengths of these resonant modes are essentially governed by the hole size (and shape), and can cause broad peaks in transmission near and beyond $\lambda_c$ except when d is very small.

The overall upshot of the above theoretical considerations is that transmission for $\lambda > \lambda_c$ in Ebbesen type conditions is actually larger than the Bethe result, and much larger than the ideal waveguide theory for a perfect conductor (by at least one order[29]) with an effective cut-off wavelength larger than $\lambda_c$ by a factor of about 2 for realistic metals[30,38]. Of course this may be offset to some extent when the waveguide and LSP related modes experience losses in the metal. Quantitative theoretical calculations of transmission spectra for different metals in a large enough range of values of h and $\lambda$ ($>\lambda_c$ >d) are expected to become available in the near future. Although not discussed here, a fair amount of work is reported on influence of the shape[40,41] of the hole on the transmission, and sensitivity to light polarization in certain cases.

Similarly, on the experimental side, definitive picture is now emerging. Some years ago, Grupp et al[42] showed that power transmitted through a circular hole (d=150nm) in a suspended silver film with h=500nm continuously decreases with $\lambda$ in the range 400nm-1000nm. Lezec and Thio[28] reported detailed measurements on single isolated holes drilled in a silver film that was deposited on fused silica substrates and topped by an index matched fluid with n=1.46. Their results, obtained in the regime 900nm>$\lambda$>$\lambda_c$ =372nm with d=150nm, h=175nm, also showed that the transmitted power decreases continuously (but nonexponentially) with $\lambda$. In their experiments on silver films suspended in air, Degiron et al[43] confirmed the decay behavior at large $\lambda$ seen by Grupp et al[42] and Lezec and Thio[28]. In addition, Degiron et al[43] found a broad peak in transmitted power at about 700nm when h is small enough i.e. comparable to d (=270nm in their case). They attributed the peak to resonant LSPs on the hole rims on the incidence and exit sides. As expected, the peak diminishes as h increases. The angular distribution of transmitted light seemed to support the dipolar radiation picture. Since light emerging from a hole is generally emitted into all angles, collection of all of the transmitted light is a challenging task for the experimentalist. This poses a problem for obtaining accurate quantitative data on the transmittance (defined as ratio of transmitted and incident power on the hole). Some estimates of the transmittance are nonetheless already available. For example, the experiments of Lezec and Thio[28] found the transmittance to be as large as 0.75 (0.25) at λ= 600nm (850nm). This was later claimed to show a good match with numerical calculations for a realistic metal[30]. Extensive measurements of transmittance in the wavelength range 450nm-850nm have been reported recently by Przybilla et al[44] for an isolated circular hole in gold with h=295 nm, deposited on glass, for different values of d ranging from 150nm-300nm. For d=300nm, the transmittance is about 27% (18%) at λ= 600nm (800nm), which decreases as d becomes smaller. Further measurements of the transmittance for holes of different geometries and materials, and their quantitative comparison with different theories, should lead to a more complete picture.

Large transmission across nm size single aperture is of much significance for those applications where large optical throughput for such apertures is imperative. High density optical data storage, nanometric microscopy and spectroscopy are some of the examples. Apart from large throughput, it is necessary to have a much more collimated light beam emerging from the nanoaperture than determined by diffraction. Patterning the metal surface around the aperture, in the form of periodic dents[42], circular grooves[45] or even finite hole arrays[44] has been found useful in this (Fig.6). It is believed that SPPs play a crucial role there, as noted below. One may recall that a metal coated tapered glass fiber tip with a subwavelength opening is usually employed in near field microscopy. Since the throughput of this waveguide is usually very small, techniques to pattern the metal coating, and thereby exploit coupling to SPPs to brighten the tip have been reported recently[46]. Previously, a patterned metal film with a hole placed on the exit face of an untapered flat fiber tip was shown to have light throughput several orders larger than the tapered fiber tip[47].

*Periodic array of holes*

How does transmission for a single isolated hole differ from that in a hole placed in a two dimensional array? In brief, the main distinguishing features are the following. Whereas the overall shape of the transmission spectra for $\lambda > \lambda_c$ is the same for both the cases[29], transmission for the array shows some maxima and minima superposed on it. At a maximum, the transmittance $T_1(N)$ for a single hole in an N x N array of holes (N>>1) may exceed unity, as shown by Przybilla[44]. This means that more light is transmitted through the hole than incident directly on it. This is the '*extraordinary* transmission'. Further, $T_1(N)$ at the maxima can be larger than that for a single isolated hole ($T_1(1)$) by a factor of a few tens. The transmission spectra have no polarization dependence for normal incidence on an array of circular holes (just like the case of a single hole). However, the wavelength positions of the maxima (and minima) show strong dependence on the angle of incidence if the light is p-polarized[24]. This is not seen for s-polarized light[24]. In comparison, the transmission in a single circular hole is essentially angle independent. One now has to understand how the holes in the array cooperate to cause these distinguishing features.

An isolated subwavelength hole acts like a scatterer of the incident plane wave. In general, the scattering can cause nonspecularly reflected waves, LSPs on the hole edges, evanescent waves in the hole cavity (for $\lambda > \lambda_c$), and a series of surface waves propagating away from the hole[28,35], a part of which can be SPPs[48]. For a single hole, the surface waves mostly dissipate away and do not contribute to transmission. However, if the hole is surrounded by a regular arrangement of grooves or dents, it is found that the transmission gets resonantly enhanced[42]. This is most likely to be due to diffraction of the SPPs excited on the patterned metal surface into the holes. Further, if the interface on the exit side also has a similar structure, its diffraction effect on the light emerging from the hole can lead to a well collimated beam[45]. One would expect that a similar effect will occur if the patterning is in fact a regular lattice of identical holes. The role of SPPs in enhancing transmission for an array of holes was first invoked by Ghaemi et al[24] who reported peaks and dips in transmission. In their measurements on an array of holes in a thick silver film on quartz substrate, Ghaemi et al found that the transmission spectra were dependent on the angle of incidence for p-polarization but not for s-polarization. Further, the transmission maxima showed dispersion similar to that of an SPP on a flat interface. The dispersion of the transmission maxima is however red shifted in λ by several nms with respect to SPP bands on a flat interface. If the transmission maxima are directly associated with SPPs on the individual perforated interfaces, one has to conclude that the large shift of the SPP dispersion relation from that for a flat metal/dielectric interface is a consequence of the interface having a lattice of subwavelength apertures. It appears that this is not fully supported by a rigorous theory as yet. (See, however, Ref. 34).

Martin-Moreno et al[31] proposed that when h is small enough, transmission maxima occur when the incident light excites a 'SPP molecule' formed because of coupling between SPPs on the two interfaces on the top and bottom sides of the perforated metal film in coincidence with a vertical Fabry-Perot type resonance of the dominant evanescent mode in the hole. The difference between the SPP relations for a flat interface and perforated interface is taken to be a combined result of the SPP coupling and the Fabry-Perot effect[34]. The enhancement in transmission is boosted when the SPP modes on the two interfaces are degenerate and coupled, but diminishes if the two metal-dielectric interfaces are dissimilar[32] or when h becomes large. In the SPP picture[31], as h increases, the incident light first excites an SPP on the front interface, the SPP then couples to the cavity evanescent mode, which then excites an SPP on the exit side interface. These processes occur in a serial fashion. In the above model, the main role of the hole array is to supply the missing momentum for optical excitation of the SPPs.

While it is generally recognized that surface waves have an important role in the transmission process[36,49], a purely SPP based point of view is questioned by some[28,30,35,36]. It is pointed out that the theory should explain both maxima and minima occurring in transmission for $\lambda>\Lambda$ more than once. The supporters of the SPP model initially attributed[24,31] strong suppression of transmission seen near $\lambda\sim\Lambda$ to an SPP unrelated phenomenon of the so-called Wood-Rayleigh (WR) anomaly. In WR anomaly, first observed by Wood[50] for one-dimensional reflection gratings and explained by Rayleigh[51], a sudden change in reflected (or transmitted) intensity occurs due to redistribution of p-polarized light energy among various diffraction orders when one of them becomes tangential to the interface as $\lambda$ increases, and vanishes with further increase in $\lambda$. This happens at $\lambda=\Lambda/m$, m=1,2..., for normal incidence (Fig. 7). The many minima seen experimentally[23] for $\lambda>\Lambda$ are not explained by the WR anomalies. (As regards Wood's anomalies, we note for completeness that a rapid variation in the diffracted intensity, occurring beyond the WR anomaly, when the evanescent diffraction order couples to a grating surface resonance (e.g. a waveguide or SPP) is another type of Wood anomaly. See Ref. 52 for further details.). The peaks and dips in transmission in the visible/IR are also obtained in experiments[29,57-59] and numerical calculations[60,61] for perforated films of materials like chromium, tungsten and nickel (which are not quite plasmonic in the visible/IR), as well as of nonmetals like amorphous silicon[28]. The dip-peak features in the non plasmonic materials are weaker compared to metals like Au and Ag, but that is expected in view of the larger penetration depth and absorption of light in these materials. Further, the extrema in transmission are seen[31,58] even in regimes where the metals are nearly perfect conductors so that no SPPs exist. The above factors raised doubts about the soundness of the SPP resonance theory. The following simple picture is proposed as an alternative[28].

Scattering of light incident normally on each hole of the array gives rise to a composite of several evanescent surface waves (including SPPs which may dominate in some cases, but not all) propagating away from the hole along the surface (Fig.8). When the composite wave arrives at a next hole, it interferes with light directly incident on that hole. Light diffracted into the hole is proportional to the interference signal. Similarly, light on the exit side of the hole launches a composite surface wave on the corresponding surface. That wave in turn scatters at another hole, interfering with light directly emerging from it. Depending upon the relative phase of the interfering waves, enhancement or suppression in the transmission results. The role of interference seems to be confirmed in theoretical simulations[59] and Young type two-slit experiments[60]. For rows of holes separated by $\Lambda$, Lezec and Thio[28] estimated that transmission maxima occur at $\lambda_m = n_{eff}\Lambda/(m-\phi/2\pi)$, where m=1,2…, $n_{eff}$ = effective refractive index experienced by the composite evanescent wave launched between the rows, and $\phi$ is its phase relative to incident wave.

The SPP based theory relies on coupling of incident light to SPPs via phase matching afforded by the periodicity of the hole lattice. SPP damping however limits its propagation length which could be much smaller than the array size. Early experiments of Grupp et al[42] had already suggested that SPP enhanced transmission phenomena may be rather local, involving neighbors within only a few periods of the hole lattice. Recent studies show that enhanced transmission occurs even in arrays of holes lacking strict translational symmetry[61]. It was suggested that light-SPP coupling is then facilitated by long range order of the array. No SPP related transmission resonances are expected for a random array according to one theory[62]. In a subsequent experiment[63], enhanced transmission in holes milled in Ag film (h=250nm) on silica substrate was found even for a random structure lacking any translational invariance and long range order. This is taken as a support for the interference model, in which light diffracted from SPPs generated by scattering of incident light at each hole is $\pi/2$ phase shifted with respect to light incident directly into a hole, giving constructive interference. According to this model, maxima may occur even with only a local, short range order.

Many claims of the interference model still remain to be fully justified. There is the need for more quantitative analysis to explain phenomena like the SPP like dispersion shown by the transmission maxima for oblique angles, and that too only for p-polarized light. As noted earlier, transmission spectra for nearly perfect conductors and some non-metallic materials also show maxima and minima. Since these materials do not support SPPs, this indicates that effects other than SPPs exist. At the same time, if these materials are substituted by plasmonic metals like Ag or Au, there is a large enhancement in the peaks, suggesting that SPPs and LSPs certainly have a role. It may also be mentioned that for p-polarization, a perfect conductor with a 2-dimensional array of holes with $\lambda >> \Lambda > d$ effectively acts like a plasmonic metamaterial, supporting a surface wave with SPP like behavior. For a square array of square holes (hole side=a, hole-hole distance=$\Lambda$) in a perfect conductor, it was shown[64] that the effective dielectric response has a plasmon type form, namely $\varepsilon_{eff} = \varepsilon_{pc}(1- \omega_{pc}^2/\omega^2)$ with the plasmon frequency $\omega_{pc} = \pi c/[a(\varepsilon_h)^{1/2}]$ where $\varepsilon_h$ is the permittivity of the dielectric material filling the holes, c=light velocity in vacuum, and $\varepsilon_{pc}=(\varepsilon_h/8)(\pi\Lambda/a)$. Here $\omega_{pc}$ is in fact the hole waveguide cut off frequency. It is interesting that although a flat surface of a perfect conductor does not support bound states, a structured surface with an array of narrow holes on a perfect conductor can also support bound states in the form of 'spoof surface plasmons' with properties that can be engineered by the hole geometry. This result reinforces the SPP based arguments for transmission maxima.

On the whole, it seems that while the diffraction and interference effects lead to peaks and dips in transmission for hole arrays in most materials, the effects get 'extraordinarily' accentuated whenever SPPs and LSPs can be excited on the perforated metal-dielectric interfaces.

**III Sub wavelength slits in an opaque metal film:**

Study of optical properties of two-dimensional array of holes can get complicated because of the influence of different shapes of holes and their lattice arrangements, especially for theoretical simulations. It is simpler to consider a one-dimensional periodic array of long, narrow rectangular slits etched in a metallic opaque film (of thickness h) supported on a dielectric substrate. Such a lamellar grating is shown in Fig.9 (a). The plane of incidence is taken to be perpendicular to the slit lines. The walls of the metallic slit and the top and bottom metal-dielectric interfaces act like a rectangular waveguide cavity[1]. In that case, a narrow slit with $\lambda>2W$ (W=slit width) supports only evanescent light modes if the light is TE i.e. s-polarized. However, if the light is TM i.e. p- polarized, a subwavelength slit always supports at least one propagating mode even when $\lambda>2W$, in addition to evanescent modes. This is a very important distinguishing feature for optical transmission across subwavelength slits and holes. It may be recalled that circular holes of diameter d do not support any propagating modes if $\lambda>\lambda_c\sim1.7d$.

Extraordinary transmission of light across an array of subwavelength metallic slits with $\lambda>\Lambda>2W$ was first analyzed by Porto et al[65] ($\Lambda$=period of the 1-D grating). As in the case of hole array, transmission and reflection of light incident on the slit array can be described by matching solutions to Maxwell's equations in the three regions: dielectric substrate, perforated metal slab and dielectric superstrate. For p-polarized light with wavelength $\lambda>2W$, coupling of incident light only with the fundamental propagating Bloch mode of the slit array may be retained to a good approximation (except when h is very small when evanescent modes may contribute). This leads to a Fabry Perot type analytical formula[65-67] for zero order transmittance for the slit array. For a symmetric slit array with identical metal-dielectric interfaces on the two sides, the transmittance is given by $T=|\tau^2\exp(i\phi_\beta(h))/[1-\rho^2\exp(2i\phi_\beta(h))]|^2$ where $\tau$ is the coefficient of coupling of the incident wave with the dominant propagating mode of the slit and $\rho$ the coefficient of reflection of the mode at the interfaces (Fig.9(b)). The phase $\phi_\beta(h)$ (=$\beta h$) arises due to propagation of the mode in the slit, with a propagation constant $\beta$ ($\cong 2\pi/\lambda$). Transmission peaks are obtained at zeros of the denominator of transmittance T. Writing $\rho$ as $|\rho|\exp[i\phi_r(\theta)]$, this happens when $|\rho| \sim 1$ and $\phi=\phi_\beta(h)+\phi_r(\theta)=m\pi$, m=integer[66] i.e. when $\lambda \cong 2h/[m-\phi_r(\theta)/\pi]$. Thus, transmission maxima for slit arrays occurring at $\lambda > \Lambda$ arise due to excitation of slit cavity resonances i.e. cavity modes (CM)[61]. Since CM related peak in transmission is essentially related to single slit effects, its position in wavelength is not sensitive to $\Lambda$[68]. On the other hand, SPP related features are expected to shift with $\Lambda$. In contrast, the CMs shift with a change in h, but the SPP related effects do not. We may note here that the wavelength of transmission peaks for hole arrays is more or less h independent but it does depend on $\Lambda$. Further, the transmission peaks for hole arrays decay exponentially[69] with h, but not in the case of slit arrays, where many new peaks emerge as h increases.

The CMs are nearly $\theta$ independent and show blue shift towards $\Lambda$ as h is reduced. At small enough h (and certain higher critical values of h), a very narrow transmission peak is seen near $\lambda \sim \Lambda$ for $\theta = 0^0$. Its wavelength is $\theta$ (and $\Lambda$) dependent and shows a dispersion that is in close proximity with the dispersion of SPP on a flat metal-dielectric interface. Porto et al[65] proposed that the narrow peak near $\lambda \sim \Lambda$ for $\theta = 0^0$ is not related to CMs but is caused by excitation of coupled horizontal SPPs on the two metal-dielectric interfaces on the slit array, in a manner similar to the transmission maxima for a hole array. How does a transmission peak seen for small h at $\lambda \sim \Lambda$ for $\theta = 0^0$, and attributed by Porto et al[65] to coupled SPPs, smoothly evolve into a CM related peak at a longer wavelength when h is made larger? Crouse and Keshavareddy[70] proposed that the transmission through the slit arrays is mediated by a hybrid mode composed of CM and SPP components, the composition changing with h. Cao and Lallane[67] argued that the θ dependent peak is not caused by SPPs but is actually related to WR anomaly. Cao and Lalanne further found that the transmission in fact vanishes at $\lambda=\lambda_0=(\Lambda/m)n_{SP}$ for $\theta\sim0^0$ (m=1,2…, $n_{SP}$ = Re$\{[\varepsilon_d\varepsilon_m/(\varepsilon_d+\varepsilon_m)]^{1/2}\}$ $\cong$1.03 at 650 nm for Ag/air interface), where the parallel momentum of the incident wave matches the SPP momentum on a flat metal-dielectric interface modulo the grating wave vector (=$2\pi/\Lambda$). At $\lambda_0$, the SPP wavelength $\lambda_{SP}$ equals $\Lambda/m$, m=1,2… for $\theta\sim0^0$. An intuitive explanation for vanishing transmission at $\lambda=\lambda_0$ for $\theta\sim0^0$ was provided by Lalanne et al[71]. It was explained that, in general, a solution of the Maxwell's equations exists for a flat interface at $\lambda=\lambda_0$ such that the field amplitude has zeroes occurring with a period $\Lambda$. The solution is valid even for an array of perforations with a period $\Lambda$ so that the field becomes null on the apertures. This means that the incident light in effect does not see the perforations and thus there is no transmission. At $\lambda=\lambda_0$, the coupling of the incident wave with the slit mode in fact vanishes[72], i.e. $\tau$ =0, implying vanishing transmission. On the other hand, the SPP mode on an individual perforated interface is identified in the Fabry-Perot picture with a 'pole' of $\rho$[66] where |ρ| ~1 and $\phi_r(\theta)\sim\pi$. For small enough h, this leads to vanishing of the denominator in T, giving a maximum in transmission. This can be shown to happen near $\lambda\sim\Lambda$ for $\theta=0^0$. Generally, |τ|, |ρ| and $\phi_r(\theta)$ vary very rapidly near $\lambda\sim\Lambda$. It turns out that the zero of |τ| and pole of |ρ| (and causing vanishing transmission and a peak in transmission, respectively) occur in close proximity of each other in $\lambda$[64,68]. The net effect could be that the transmission maximum related to the pole of ρ (and hence to SPP on the slit array) gets suppressed. For the zero and peak in transmission to be clearly distinguished, the SPP dispersion on the metal surface perforated with slits has to be well separated from that on a flat interface. How large the effect of the perforation, if any, is on the SPP dispersion is not rigorously established yet.

In the case of slit arrays, the importance of CMs in giving peaks in transmission is now well recognized. However, many other issues have been under continued discussion, such as the role of SPPs in determining the transmission extrema, modifications in SPP dispersion on flat metal-dielectric interfaces, if any, by the perforations, relationship of the peaks and dips in transmission near $\lambda \sim \Lambda$ with the zeros and poles of $\rho$ and $\tau$ and with the two types of Wood anomalies (WR and SPP), and so on. Most of the early theoretical and experimental studies were made in the infrared region ($\lambda > 1\mu m$), where the WR and SPP anomalies occur close to each other in $\lambda$, making their identification with the dips and peaks in transmission rather uncertain. Rigorous numerical calculations[72-74] of the optical fields obtained recently for normal and oblique incidence of p-polarized light on lamellar slit gratings have now confirmed that the transmission nearly vanishes at $\lambda=\lambda_0$ whereas a peak near $\lambda \sim \Lambda$ for $\theta = 0^0$ is related to WR anomaly.

In comparison with the large number of theoretical studies performed, relatively few definitive experimental results on metallic slit gratings are available for comparing with the theories. Many of the early experiments were interpreted[75,76] to comply with the SPP model. Recently, we reported extensive transmission and reflection measurements[77] in the visible/IR on 200nm thick Au film, deposited on a periodic array of grooves etched in quartz substrate with a depth of 600nm and width of $\Lambda/4$, for 740nm>$\Lambda$>600nm. For this structure, the peaks caused by CM and WR, as well as the WR and SPP features near $\lambda \sim \Lambda$ for $\theta = 0^0$ (namely poles of $\tau$ and $\rho$ respectively) are expected to be well separated in $\lambda$. The profile of the deep trench grating used in this study (Fig.10) suggests that propagation of SPPs on the metal-quartz interface would be impeded. In contrast, propagating SPPs on the metal-air interface can bridge the narrow gap of the groove, just as in the case of a lamellar grating (Fig. 9). Indeed, transmission measurements reveal the WR anomalies on the grating at the metal-quartz interface but no signature is seen for excitation of SPPs on that interface. Similar situations where SPPs on the metal-substrate are precluded were investigated by Crouse and Keshavareddy[70]. The role of SPP and WR anomaly on the *metal-air interface* of the grating in giving extrema in transmission thus can be well clarified. The experiments clearly vindicate the theoretical notion that the transmission vanishes when the light wavelength matches $\lambda_0$ and shows a peak at the WR anomaly[67,72-74]. The results suggest that SPP dispersion for this grating is not very different from that on a flat interface, implying thereby that the zero of $\tau$ and pole of $\rho$ nearly overlap. The loci of the dips in transmission show good agreement with theoretical calculations[78] of SPP dispersion based on S matrix theory with realistic $\varepsilon_m$ used for Au, going beyond the Drude model. The spectral widths of the dips associated with SPPs are consistent with SPP lifetime deduced from femtosecond time domain measurements[79]. A strong dependence of the optical properties on polarization and polar and azimuthal angles of the incident light is found[80]. The results additionally confirm that the dips in transmission are related to excitation of SPPs on the metal-air interface. An interesting issue is: how different are the transmission/ reflection spectra when light is incident from the top and substrate side. The measurements[77,80] show perfect reciprocity for transmission, in agreement with a theoretical proof of reciprocity in transmission given earlier[81]. However, reciprocity is not seen in reflection. As shown in Ref. 82, this property is generally expected for one dimensional stratified media with broken inversion symmetry, together with absorption. To explain the dips seen in transmission, a simple picture based on interference of light propagating directly through the slit, and that diffracted from the surface wave (here, SPP) was proposed. It was suggested[77,79,80] that a dip in transmission occurs at a wavelength corresponding to excitation of an SPP on the metal-air interface because the SPP-diffracted and directly-transmitted field into the slit interfere with opposite phase. Because of this, a Fano lineshape signal generally expected for interfering resonant and continuum states becomes a symmetric dip in the present case. Recently, Pacifici et al[83] studied transmission in a few-slit sample. They deduce a formula for transmission based on a picture[59,60] in which SPP generated at a slit by the incident light propagates to a neighboring slit where it diffracts into the slit cavity. The field at the cavity entrance is a superposition of the incident field and the SPP diffracted field. The superposed field launched into the slits undergoes multiple reflections at the exit and entrance sides of the slits, as in Fabry-Perot cavity. The extrema are governed by the phase $k_{SP}\Lambda$ ($\Lambda$=slit separation) accumulated by the SPP to travel from one slit to the next plus the phase $\varphi$ arising due to conversion of incident light into SPP at slit 1 and the reverse conversion at slit 2. Since the minima in transmission are seen[83] exactly at $\Lambda=m\lambda_{SP}$, m=1,2,3.., $\varphi$ is deduced to be $\pi$. (Interestingly, in another work, Pacifici et al[63] deduced $\varphi=\pi/2$ for holes. The difference was attributed to dependence of $\varphi$ on the aperture shape, with the long narrow slit and subwavelength circular hole respectively acting like a line and a point dipole). The minima are caused by destructive interference between incident waves and the diffracted SPP on a slit. This model attempts to incorporate many of the relevant physical concepts like CM, SPP, diffraction and interference in an intuitive manner.

Finally, one may mention that a transmission mechanism similar to that in hole arrays may occur for slits in a certain situation. As already discussed, for slit arrays, neither a propagating CM nor a SPP mode exists for s-polarization. Yet, transmission maxima can occur even for s-polarized incident light if it can excite a surface mode, such as a guided wave supported by a thin overlaying dielectric film on top of the slit grating[49]. This situation is similar to the case of a SPP mode excited on a hole array (which too has no propagating mode in the aperture). For a symmetric structure with a similar dielectric film on both sides, a transmission maximum may occur via tunnel-coupling of the surface modes through the slits. This supports the notion[36,49] that, in general, transmission maxima may occur when any type of coupled surface wave resonance (including but not necessarily SPP) is excited - for hole as well as slit arrays

## IV Concluding remarks

Observation of unexpectedly large optical transmission in subwavelength openings in opaque metal slabs has caused tremendous excitement because of its implication for several novel applications in nanophotonics. A very large number of reports on this topic have appeared in the literature in the past decade or so. Much effort has been expended for understanding the physics of the observed phenomena but the consensus on this is still emerging. It is realized that there are important differences in light energy transfer mechanisms for an array of subwavelength holes and slits. Many complex effects involving surface waves, diffraction, interference and localized modes in individual apertures need to be taken into account together to fully understand the peculiar features seen in the optical transmission spectra. In this non exhaustive, limited review, we focused primarily on the status of this effort. The various exciting applications proposed and demonstrated were not covered. It is expected that with continued work using sophisticated experiments and numerical simulations, our understanding of the physics of the 'extraordinary optical transmission' will be complete in the near future. One hopes that this will lead to further breakthroughs in nanophotonics applications.

*(MS Completed in Oct. 2009, to appear in Current Science, J. of Ind. Acad .Science)*

**Figure captions**

1.  (a) Transverse magnetic (TM) i.e. p-polarized light, with the electric vector **E** in the plane of incidence, is necessary for coupling with surface plasmon polaritons (SPP) on a smooth metal-dielectric interface, but this is not sufficient. Light incident from the dielectric can not couple to SPPs on the interface because the wave vector component of light along the interface remains smaller than that of the SPP even for grazing angle incidence ($\theta=\pi/2$). This is further clarified in (b) which shows the SPP dispersion on a metal-air interface schematically along with light lines. Note that the light lines always lie to the left of the SPP dispersion.

2.  (a) Light incident on an interface between two dielectric materials $D_{II}$ and $D_I$ with refractive indices $n_{II}$ and $n_I$ ($n_{II}>n_I$) from $D_{II}$ undergoes total internal reflection at a certain angle of incidence ($\theta_{TIR}$). The refracted wave propagating along the interface becomes evanescent for $\theta >\theta_{TIR}$, its field decaying into $D_I$. (b) Same as (a) except that now a thin enough metal film (M) separates $D_{II}$ and $D_I$. This is the Kretschmann geometry shown in a semi cylindrical prism configuration.. The refracted evanescent wave tunneling across the film can satisfy phase matching with a surface plasmon polariton (SPP) on the metal- $D_I$ interface and couple to the SPP if the light is p - polarized. (c) If the $D_I$ layer in (a) is thin enough and topped by a metal layer (M), the tail of the evanescent wave refracting at the $D_{II}$ - $D_I$ interface can couple to the SPP on the $D_I$ –metal interface, if the light is p-polarized. This is the Otto prism geometry.

3.  A scheme to excite coupled surface plasmon polaritons on the interfaces of a thin metal film (M) with a dielectric material $D_I$ on its two sides is shown using two back-to back Otto type configurations. The dielectric $D_{II}$ is as explained in Fig.2. This arrangement can lead to large transmission across an otherwise semi-opaque metal film[10].

4. If a smooth metal (M) -air interface is replaced by one dimensional periodic corrugation of period $\Lambda$, Bragg scattering by the grating provides for the mismatch between the surface plasmon polarion (SPP) wave vector ($k_{SP}$) and the interface (x) component of the wave vector of incident light ($k_{ph}\sin\theta$), thus establishing phase matching at a certain angle of incidence $\theta$. G is the grating Bragg vector. The incident light is TM-polarized, with the electric (**E**) and the magnetic (**H**) vectors of the light as shown.

5. Opaque metallic films with an array of subwavelength holes and a single isolated hole are shown. It is meaningful to compare per hole transmission of the array with that of a single isolated hole for investigating the effect of the array.

6. A schematic drawing is shown to indicate how the effect of surface wave diffraction and interference caused by a periodic structure of grooves or dents surrounding a single metallic subwavelength hole can contribute to enhance and collimate light transmission through the hole.

7. A propagating diffracted order (dashed arrow) for corrugated metal (M) -air interface is shown for normal incidence, with the x component of its wave vector ($k_x$) given as $k_{ph}\sin\theta_d = \pm mG$, m=1,2…according to the grating equation. $G=2\pi/\Lambda$ is the grating Bragg vector As the light wavelength ($\lambda$) (and therefore the angle of diffraction $\theta_d$) increases, a diffraction order can become tangential to the grating such that $\sin\theta_d \rightarrow 1$. $k_{ph} = (2\pi/\lambda)$ then equals mG, m=1,2... just before the diffraction order vanishes. This is the condition of Wood-Rayleigh anomaly. A further increase in $\lambda$ makes the normal component ($k_z$) of its wave vector (=$k_{ph} \cos\theta_d$) imaginary and the vanishing diffraction order becomes evanescent. Since $k_{ph} = (2\pi/\lambda) = [k_x^2+k_z^2]^{1/2}$, $k_x > k_{ph}$. The evanescent order can then phase match and couple to a surface plasmon polariton resonance. This is a kind of 'Wood anomaly'.

8. Diffraction and interference effects of an array of subwavelength holes with $\lambda > \Lambda$ on incident and transmitted light can lead to collimation and enhancement or suppression of zero order transmission depending upon constructive or destructive phase relationship of directly incident and diffracted light components at the hole entrance and exit. This is shown here schematically.

9. (a) A lamellar structure of an array of slits of subwavelength widths in a metallic slab on a dielectric substrate is shown schematically. The plane of incidence is perpendicular to the slits. The incident light is TM (or TE) i.e. p (or s) -polarized if the light magnetic (or electric) vector is perpendicular to the plane of incidence. The double sided arrows show surface plasmon polaritons propagating on metal-air and metal-substrate interfaces. (b) A simple Fabry-Perot model is often used in the literature[65-67] to describe transmission of light across the slit array in terms the coefficient of coupling of the incident wave with the dominant propagating mode of the slit ($\tau$) and the coefficient of reflection of the mode at the interfaces ($\rho$).

10. A grating structure of deep grooves etched in quartz substrate with a metal film deposited on top is shown schematically. The double sided arrows indicate that propagating surface plasmon polaritons can occur on the metal-air interface but not likely on the metal-substrate interface.

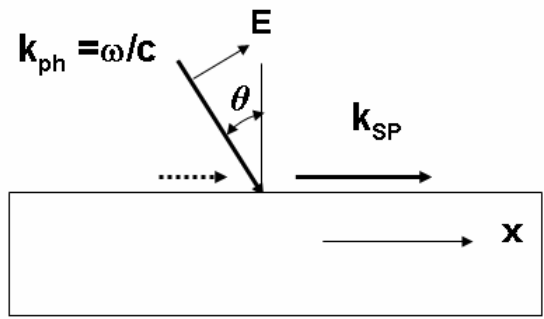

(a) $k_{SP} > k_x = k_{ph} \sin\theta$

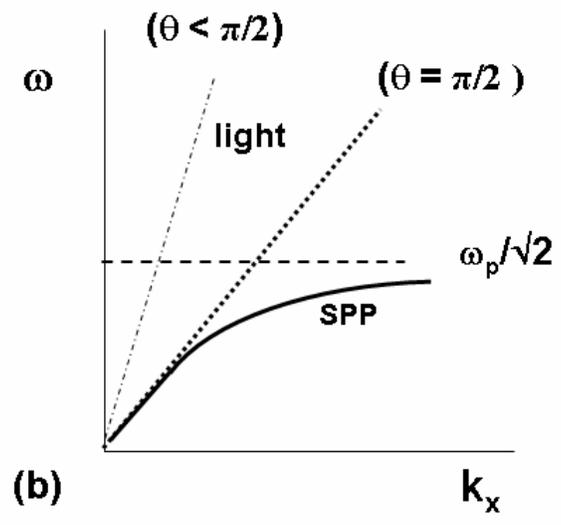

(b)

Fig.1

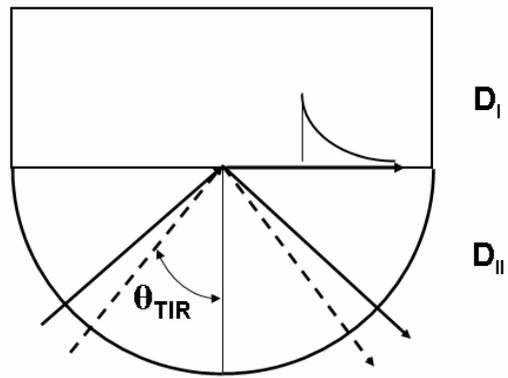

(a)

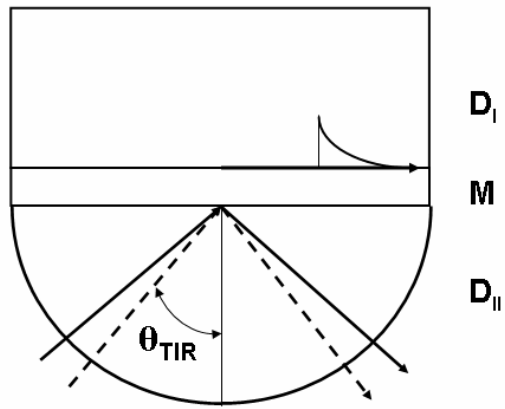

(b)

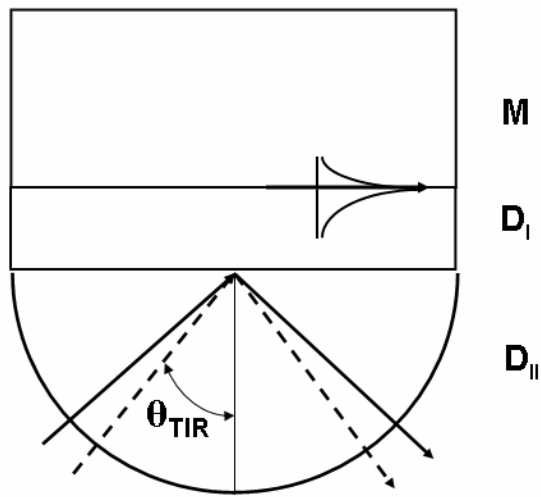

(c)

Fig.2

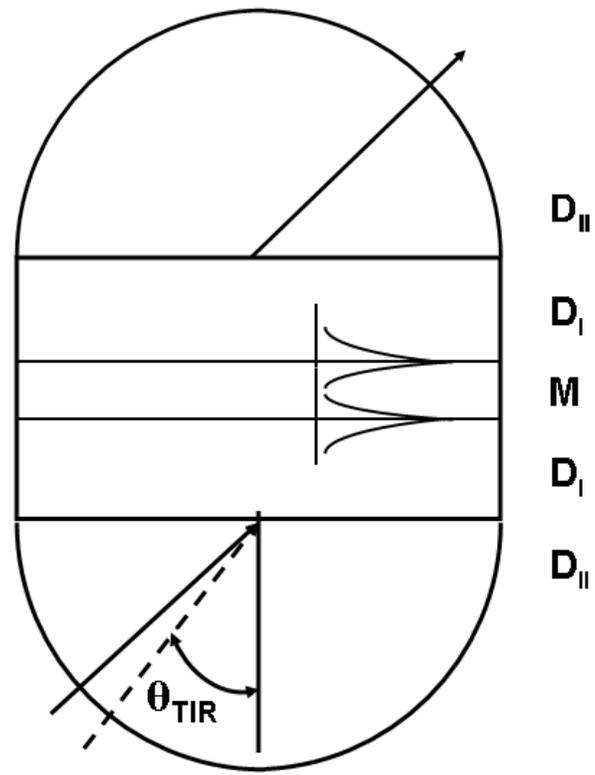

Fig.3

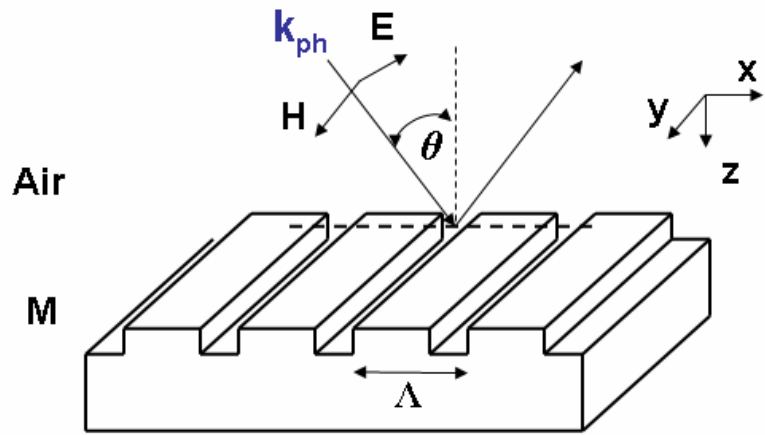

Fig.4

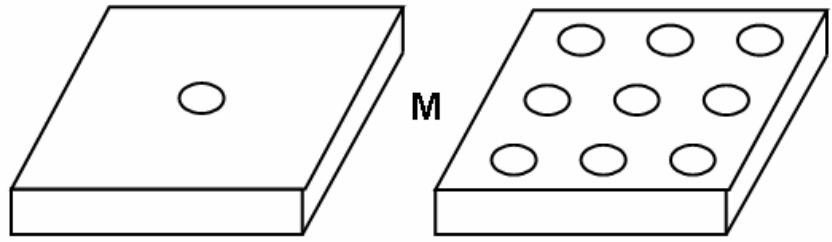

Fig.5

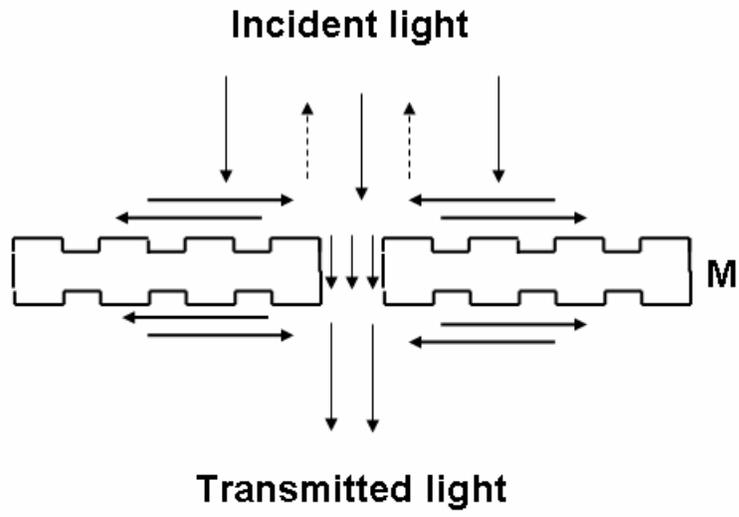

Fig.6

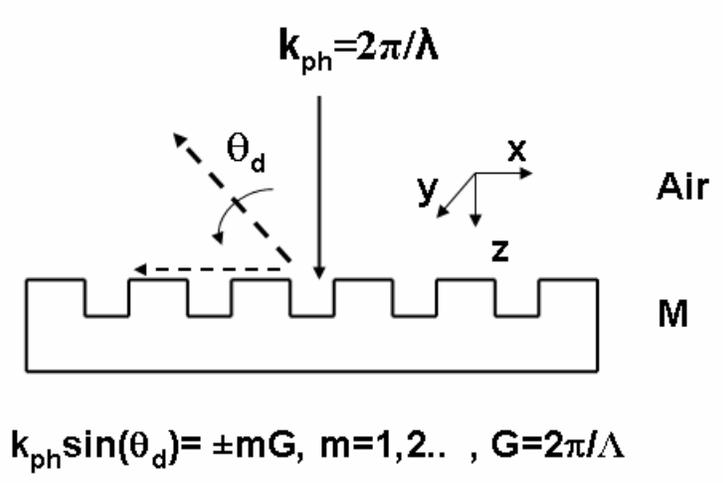

Fig.7

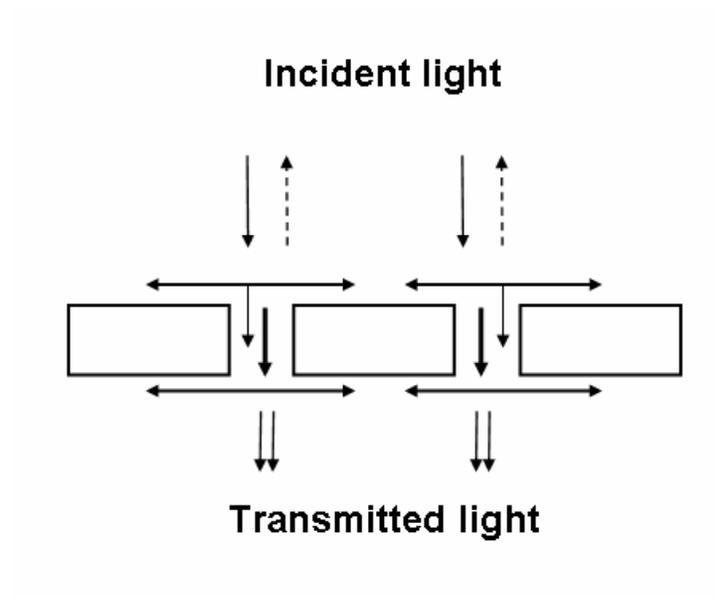

Fig.8

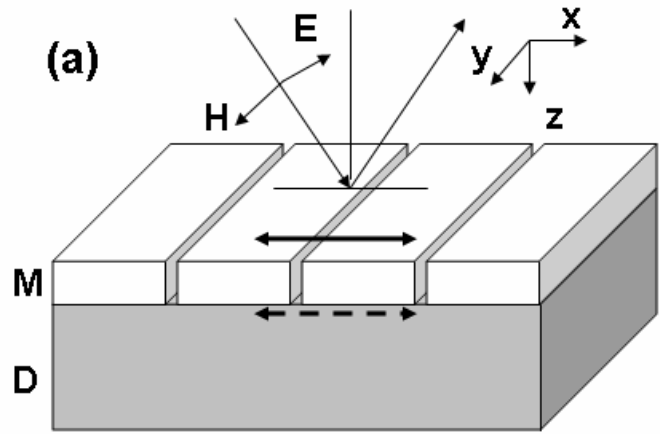

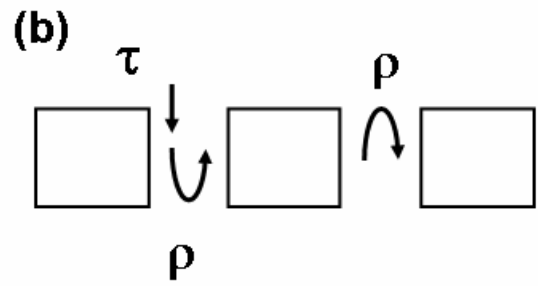

Fig.9

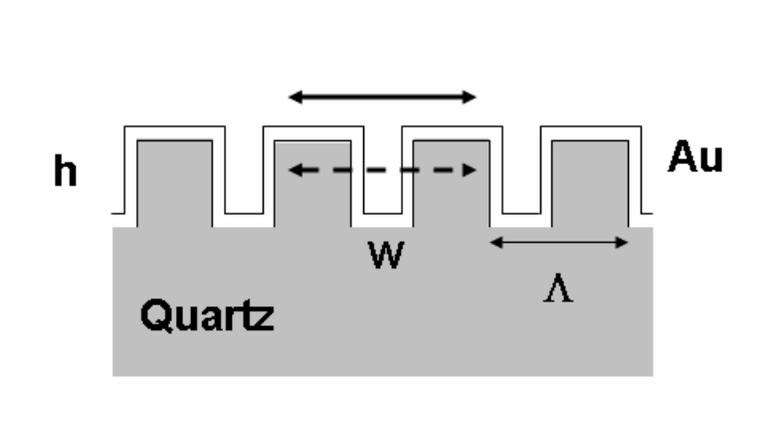

Fig.10